\documentclass[12pt,a4paper]{article}

\renewcommand{\theequation}{\thesection.\arabic{equation}}

\newcounter{appendix}
\newcommand{\newappendix}[1]{\vspace{10mm}\pagebreak[3]
\addtocounter{appendix}{1}
\renewcommand{\theequation}{\Alph{appendix}.\arabic{equation}}
\setcounter{equation}{0}
\begin{flushleft}{\Large\bf Appendix \Alph{appendix} #1}
\end{flushleft}\nopagebreak\medskip\nopagebreak}

\def\be{\begin{equation}}
\def\ee{\end{equation}}
\def\bea{\begin{eqnarray}}
\def\eea{\end{eqnarray}}

\usepackage{amsfonts}
\begin{document}
\begin{flushright}
Ref. SISSA 86/2000/FM
\end{flushright}
\vspace{1.truecm}
\begin{center}

{\LARGE{\bf{The spin 1/2 Calogero-Gaudin System }}}
\medskip
{\LARGE{\bf{and its $q-$Deformation}}}

\vskip1cm

{\large{\sc Fabio Musso}}\vspace{.2cm}\\
{\it S.I.S.S.A.}\\
{\it Via Beirut 2/4, I-34013 Trieste, Italy}
\vskip0.5cm

{\large{\sc Orlando Ragnisco}}\vspace{.2cm}\\
{\it  Dipartimento di Fisica,Universit\`a di Roma  TRE,}\\ 
{\it Via Vasca Navale 84, 00146 Roma, Italy} \\
{\it  I.N.F.N. - Sezione di Roma TRE, Roma, Italy}
\vspace{.3cm}
\end{center}
\bigskip

\begin{abstract}
 The  spin 1/2 Calogero-Gaudin system 
and its $q$-deformation are exactly solved: a complete set of commuting
observables is diagonalized, and the corresponding eigenvectors and
eigenvalues are explicitly calculated. The method of solution is purely
algebraic and relies on the co-algebra symmetry  of the model. 
\end{abstract}
\medskip \centerline{$Running$  $Title$: Spin 1/2 CG System and its
$q-$Deformation}

\medskip
\centerline{P.A.C.S.: 03.65 Fd, 02.20 Sv}
\vfill
\eject

\section{Introduction}

The role of co-algebra symmetry for constructing integrable classical and
quantum mechanical systems has been elucidated in a number of recent
papers (see for instance \cite{Rag},\cite{Ragn}, \cite{Herra}). In particular,
this tool has been used in \cite{Mussorag} to solve the quantum Calogero-Gaudin
system \cite{C&vD},  i.e. a quantum-mechanical model built-up out of the Lie
algebra $sl(2)$ and its $q$-deformation in an infinite-dimensional
representation. The purely algebraic nature of the approach followed in
\cite{Mussorag} was a clear indication  that the same task could have been
accomplished in any finite-dimensional representation as well, both in the
undeformed and in the $q$-deformed case. In the present paper, we are going to
present the exact solution of the system in the simplest finite-dimensional
representation, namely the spin 1/2 one. 

\noindent
We notice that the
Hamiltonian of the system we are going to study coincides - in the undeformed
case - with that of the celebrated Gaudin Magnet, which has been the subject
of a number of remarkable papers \cite{Gaudin1},\cite{Gaudin2} \cite{Sklianin},
and is a prototype example of an integrable spin chain with long range
interaction. So, one might argue that, at least for the undeformed system, we
are just proposing an alternative approach to derive well known results:
however, this is not the case - as already mentioned in \cite{Mussorag}-
because we are indeed diagonalizing a family of commuting observables which is
independent on  the one usually considered in the literature (see for
instance \cite{Sklianin}); the existence of two independent complete sets of
commuting observables is due to the fact that the system under
scrutiny is (maximally) super-integrable. It is remarkable that  when
diagonalizing our family of commuting observables, which is the sequence of
coproducts of the original Casimir operator, no spectral parameter enters into
the game, and correspondingly no (algebraic) Bethe equations have to be
solved. Moreover, we want to stress that this very family of observables,
being constructed through the coproduct, can be naturally lifted to the
$q$-deformed case, preserving the commutativity property.

\noindent 
The organization of the paper is quite simple. In Section 2, we tersely
recall the results derived in \cite{Mussorag} and then discuss the spin 1/2
undeformed case. In Section 3, the results for the $q$-deformed case are
given.In Section 4 we draw a few concluding remarks and outline some possibly
interesting open problems. To speed up the presentation, the heaviest part of
the calculations is confined to Appendix A; in Appendix B
we give the explicit form of the basis spanning the "Bethe sea"
for a number of sites $N=5$.

\section{The undeformed spin 1/2 case}

In  \cite{Mussorag} we have studied the quantum system 
characterized by the Hamiltonian
\begin{eqnarray}
&&\hat{\mathcal{H}}=\frac{1}{4}(\lambda \hat{\cal{N}}^2+ \mu r^2 
\nabla^2) \label{HC}\\
&& \hat{\cal{N}}=\sum_{j=1}^N  \hat{\cal{N}}_j=\sum_{j=1}^N x_j 
\frac{\partial}{\partial x_j} \qquad r^2=\sum_{j=1}^N x_j^2 \qquad 
\nabla^2=\sum_{j=1}^N \frac{\partial^2}{\partial x_j^2} \nonumber   
\end{eqnarray}
on the Hilbert space of the analytic functions of the variables
$x_{1}^2,x_{2}^2,\dots,x_{N}^2$.

We have shown that if one considers the infinite-dimensional 
representation of $sl(2)$ given by:
\begin{eqnarray}
\hat{X}^3&=&\frac{1}{2}(b^\dagger b+ b b^\dagger) \nonumber \\
\hat{X}^+&=&\frac{(b^\dagger)^2}{2} \label{algebra}\\ 
\hat{X}^-&=&-\frac{b^2}{2} \nonumber
\end{eqnarray}
where 
\begin{displaymath}
b^\dagger = x \qquad b = \frac{\partial}{\partial x}
\end{displaymath}
then the Hamiltonian turns out to be of the form
\begin{equation}
\hat{\mathcal{H}}=\frac{1}{4}(\frac{\lambda}{2}+\mu) \Delta^{(N)} (\hat{X}^3)^2
-\frac{\mu}{2} \Delta^{(N)}(\hat{\mathcal{C}})
+\frac{1}{32} \lambda \label{form}
\end{equation}
Here $\hat{\mathcal{C}}$ is the Casimir operator of the algebra
(\ref{algebra}) and  the operator $\Delta^{(N)}$, together with the operators 
$\Delta^{(3)},\dots, \Delta^{(N-1)}$,  
is defined through the recursive formula:
\begin{equation}
\Delta^{(i)}=(\Delta^{(i-1)} \otimes id) \Delta^{(2)} \qquad i=3,\dots,N 
\label{Deltarec} 
\end{equation}
where $\Delta^{(2)}$ denotes the coproduct associated to the 
usual Hopf-algebra structure defined on a universal enveloping 
algebra $U(g)$, namely:
$$\Delta^{(2)}(X) = X\otimes 1 + 1\otimes X \quad \forall \ X \in g$$  

The particular form (\ref{form}) ensures us that the quantum system 
described by the Hamiltonian (\ref{HC}) is completely integrable
(actually, super-integrable)
and the $N-1$ observables commuting among them and with the 
Hamiltonian are provided by the coproducts of the Casimir operator of 
the algebra (\ref{algebra}), \cite{Rag}, \cite{Ragn}. 

Note that the problem of the simultaneous diagonalization of the 
complete set of commuting observables
\begin{displaymath}
\{\hat{\mathcal{H}},\Delta^{(2)}(\hat{\mathcal{C}}),
\dots,\Delta^{(N)}(\hat{\mathcal{C}}) \} 
\end{displaymath}
is completely equivalent to the problem of simultaneous 
diagonalization of the alternative complete set of commuting observables
\begin{equation}
\{\Delta^{(N)}(\hat{X_3}),\Delta^{(2)}(\hat{\mathcal{C}}),
\dots,\Delta^{(N)}(\hat{\mathcal{C}}) \} \label{family}  
\end{equation}
In \cite{Mussorag} we solved this problem and we found the following 
expression for the eigenfunctions of (\ref{family}) \footnote{The slightly 
different notation used in \cite{Mussorag} is indeed misleading, as it does
not explicitly show the dependence of the eigenfunctions upon the whole set of
variables.}: 
\begin{eqnarray}
&&\phi(k,m_{l},s_{m_l},m_{l-1},s_{m_{l-1}},\dots,m_{1},s_{m_{1}},0,0)=\nonumber \\ &&=[\Delta^{(N)}(\hat{X}^+)]^{k-m_{l}} 
H(m_{l},s_{m_l},m_{l-1},s_{m_{l-1}},\dots,m_{1},s_{m_{1}},0,0)\label{phi}
\end{eqnarray}
where $H(m_{l},s_{m_l},\dots,0,0)$ is an 
``$s_{m_l}$-particle harmonic polynomial'', i.e. it satisfies:
\begin{equation}
\Delta^{(s_{m_l})}(X^-)H(m_{l},s_{m_l},\dots,0,0) =0 \label{harmonic}
\end{equation}
These harmonic polynomials are generated through the recursive formula:
\begin{eqnarray} 
&& H(m_{l},s_{m_l},m_{l-1},s_{m_{l-1}},\dots,m_{1},s_{m_1},0,0) = \nonumber\\ 
&& \left( \sum_{i=0}^{m_{l}-m_{l-1}} a_{i,s_{m_{l}},m_{l},m_{l-1}}  
(\hat{X}^+_{(s_{m_l})})^{m_{l}-m_{l-1}-i} 
[\Delta^{(s_{m_l}-1)}(\hat{X}^+)]^i \right) \label{second} \\
&& H(m_{l-1},s_{m_{l-1}},\dots,m_{1},s_{m_1},0,0) \nonumber \\ 
&& m_{i}=1,2,\dots \quad s_{m_{i}}=2,\dots,N  \quad
m_{i-1}<m_{i},s_{m_{i-1}}<s_{m_{i}} \nonumber\\
&& i=1,\dots,l \quad l=1,\dots,N-1 \nonumber \\
&& H(0,0)=\overbrace{| 0 \rangle \cdots | 0 \rangle}^N=const \nonumber
\end{eqnarray}  
where the constant $a_{i,s_{m_{l}},m_{l},m_{l-1}}$ must be chosen in such a way 
that (\ref{harmonic}) holds. This condition implies the following 
recurrence relation for the
coefficients $a_{i,s_{m_{l}},m_{l},m_{l-1}}$:
\begin{eqnarray}
&& 
a_{i+1}=-\frac{(m_{l}-m_{l-1}-i)[\lambda^{min}+m_{l}-m_{l-1}-i-1]}{(i+1)[(s_{m_{l}}-1) 
\lambda^{min} +i+2m_{l-1}]} a_i \label{rec2}\\
&& \qquad i=0,\dots,m_{l}-m_{l-1}-1 \nonumber  
\end{eqnarray}
where for simplicity the labels $s_{m_{l}},m_{l},m_{l-1}$ have been omitted.
Here $\lambda^{min}$ is the minimum in the spectrum of $\hat{X}^3$; in 
our case we have $\lambda^{min}=0$.   

All these results were obtained working on pure algebraic grounds so 
we expect them to hold (with the due modifications) even if we change 
the algebra representation (\ref{algebra}). As explained in the introduction,
here we are concerned  with a particular, finite-dimensional, $sl(2)$
representation,  namely the spin $1/2$ representation:
\begin{eqnarray}
X^3&=& \sigma^3 \nonumber \\
X^+&=& \sigma^+\label{algebra2}\\ 
X^-&=& \sigma^- \nonumber
\end{eqnarray}
With this choice, the $N$-th coproduct of the Casimir gives us the 
Hamiltonian of the so-called ``Gaudin magnet'':
\begin{eqnarray}
C_N=\Delta^{(N)}(C)&=&[\Delta^{(N)}(\sigma^3)]^2 + 2\{ \Delta^{(N)}(\sigma^+),
\Delta^{(N)}(\sigma^-) \}= \nonumber\\
&=&\sum_{j,k=1}^N(\sigma_j^x \sigma_k^x +\sigma_j^y \sigma_k^y +\sigma_j^z
\sigma_k^z)=H_{G} \label{HG}
\end{eqnarray}
(we will pose $C_l=\Delta^{(l)}(C)$ $l=2,\dots,N$ for shortness).
In this representation a complete set of independent commuting observables
is again provided by the operators  
\begin{equation}
\{ \Delta^{(N)} (\sigma^3), C_{2},\dots, C_{N} \} \label{family2}
\end{equation}	

We want to find common eigenvectors of this set of observables forming a basis for the
Hilbert space of the problem.     

We will choose as ``ground state'' the state having spin down in each 
site and we will denote it again with the symbol $H(0,0)$:
\begin{equation}
H(0,0)=\overbrace{|\downarrow \rangle \cdots |\downarrow \rangle}^N
\label{ground} \end{equation}
Our claim is that, exactly as in the infinite-dimensional case, the simultaneous 
eigenstates for the family (\ref{family2})
take the form:
\begin{equation}
\phi(k,m_{l},s_{m_l},\dots,0,0)
=[\Delta^{(N)} (\sigma^+)]^{k-m} H(m_{l},s_{m_l},\dots,0,0) \label{eigenstates}
\end{equation}
$H(m_{l},s_{m_l},\dots,0,0)$ being an element of the basis of the kernel of 
$\Delta^{(s_{m_{l}})} (\sigma^-)$ 
obtained through the recursive formula:
\begin{eqnarray}
&& H(m_{l},s_{m_l},\dots,0,0)= \nonumber \\
&& =\left( \sum_{i=0}^{m_{l}-m_{l-1}} \alpha_i 
(\sigma^+_{(s_{m_{l}})})^i 
[\Delta^{(s_{m_{l}}-1)} ( \sigma^+ )]^{m_{l}-m_{l-1}-i} \right) \label{kernel}\\
&& H(m_{l-1},s_{m_{l-1}},\dots,0,0) \nonumber
\end{eqnarray} 
Here $\sigma^+_{(s)}$ is a shorthand for $\overbrace{id \otimes \cdots \otimes id}^{s-1} 
\otimes \sigma^+ \otimes \overbrace{id \otimes \cdots \otimes id}^{N-s}$ and
the coefficients $\alpha_i$ are determined by the equation
\begin{displaymath}
\Delta^{(s_{m_{l}})} (\sigma^-) H(m_{l},s_{m_l},\dots,0,0)=0
\end{displaymath}
We must introduce further restrictions on the possible values of the 
constant $k,m_{l},s_{m_{l}},\dots$ due to the nilpotency of the 
operators (\ref{algebra2}).

First of all we notice that it must hold $k \geq m$; moreover since the
state  $\phi(k,m_{l},s_{m_{l}},\dots,0,0)$ has spin $2k-N$, so that we must
have  $k \leq N$ as well. This condition  can be further restricted imposing
the simmetry of the states for spin flip which implies that $k \leq N-m$.
Moreover:
\begin{equation}
\Delta^{(s_{m_{l}})} (\sigma^3) H(m_{l},s_{m_{l}},\dots,0,0)
=(2m_{l}-s_{m_{l}}) H(m_{l},s_{m_{l}},\dots,0,0) \label{spin}
\end{equation}
Using commutation relations, we have:
\begin{eqnarray*}
&& \Delta^{(s_{m_{l}})} (\sigma^3) H(m_{l},s_{m_{l}},\dots,0,0)=\\
&& =[\Delta^{(s_{m_{l}})} (\sigma^+),\Delta^{(s_{m_{l}})} (\sigma^-)]
H(m_{l},s_{m_{l}},\dots,0,0)= \\ 
&&=-\Delta^{(s_{m_{l}})} (\sigma^-)\Delta^{(s_{m_{l}})} (\sigma^+)
H(m_{l},s_{m_{l}},\dots,0,0)
\end{eqnarray*}	
Since the operator 
$\Delta^{(s_{m_{l}})} (\sigma^-)\Delta^{(s_{m_{l}})}(\sigma^+)$
is positive, it must hold:
\begin{equation}
s_{m_{l}} -	2m_{l} \geq 0 \quad i.e. \quad s_{m_{l}} \geq 2m_{l} 
\label{inequality}
\end{equation}	
Since $\sigma^+$ is nilpotent of order two, 
the index $i$ in (\ref{kernel}) can take only two values,
namely $0$ and $1$; 
we will use the following equality:
\begin{displaymath}
\Delta^{(s)} (\sigma^-)=\Delta^{(s-1)} (\sigma^-) \otimes id + \Delta^{(s-1)} (id) 
\otimes \sigma^-
\end{displaymath}
So, we have:
\begin{eqnarray*}
&&\Delta^{(s_{m_{l}})} (\sigma^-) H(m_{l},s_{m_l},\dots,0,0)=\\
&&=(\Delta^{(s_{m_{l}}-1)} (\sigma^-) \otimes id + \Delta^{(s_{m_{l}}-1)} (id)
\otimes \sigma^-) \{ \alpha_0 [\Delta^{(s_{m_{l}}-1)} 
(\sigma^+)]^{m_{l}-m_{l-1}}+  \\
&& +\alpha_1 \sigma^+_{(s_{m_{l}})} [\Delta^{(s_{m_{l}}-1)} 
(\sigma^+)]^{m_{l}-m_{l-1}-1} \} H(m_{l-1},s_{m_{l-1}},\dots,0,0)=\\
&&=\{ \alpha_0 [\Delta^{(s_{m_{l}}-1)} (\sigma^+)]^{m_{l}-m_{l-1}} 
\Delta^{(s_{m_{l}}-1)} (\sigma^-)-\\
&& -\alpha_0(m_{l}-m_{l-1})
(m_{l}-m_{l-1}-1)[\Delta^{(s_{m_{l}}-1)} 
 (\sigma^+)]^{m_{l}-m_{l-1}-1}- \\
&& -\alpha_0 (m_{l}-m_{l-1}) [\Delta^{(s_{m_{l}}-1)} 
(\sigma^+)]^{m_{l}-m_{l-1}-1} \Delta^{(s_{m_{l}}-1)} (\sigma^3)+ \\
&&+\alpha_1 \sigma^+_{(s_{m_{l}})} [\Delta^{(s_{m_{l}}-1)} 
(\sigma^+)]^{m_{l}-m_{l-1}-1} \Delta^{(s_{m_{l}}-1)} (\sigma^-)-\\
&&-\alpha_1(m_{l}-m_{l-1}-1)(m_{l}-m_{l-1}-2) \sigma^+_{(s_{m_{l}})} 
[\Delta^{(s_{m_{l}}-1)} (\sigma^+)]^{m_{l}-m_{l-1}-2}- \\ 
&& -\alpha_1 \sigma^+_{(s_{m_{l}})} [\Delta^{(s_{m_{l}}-1)} 
(\sigma^+)]^{m_{l}-m_{l-1}-2} \times \\
&& \times \Delta^{(s_{m_{l}}-1)} (\sigma^3)
+\alpha_1 \sigma^+_{(s_{m_{l}})} 
[\Delta^{(s_{m_{l}}-1)}\sigma^+]^{m_{l}-m_{l-1}-1}\sigma^-_{(s_{m_{l}})}- \\
&&-\alpha_{1}[\Delta^{(s_{m_{l}}-1)}
(\sigma^+)]^{m_{l}-m_{l-1}-1}\sigma^3_{s_{m}} \}
H(m_{l-1},s_{m_{l-1}},\dots,0,0)   \end{eqnarray*}
Here we used the relations:
\begin{eqnarray*}
&&[\sigma^-,(\sigma^+)^m]=-m(m-1) (\sigma^+)^{m-1}-m(\sigma^+)^{m-1} \sigma^3\\
&&[\sigma^3,(\sigma^+)^m]=2m(\sigma^+)^m
\end{eqnarray*}
We note that for hypotesis it holds 
$\Delta^{(s_{m_{l}}-1)} (\sigma^-) H(m_{l-1},s_{m_{l-1}},\dots,0,0)=0$. 
Moreover, since $s_{m_{l-1}}<s_{m_{l}}$, we have as well 
$\sigma^-_{s_{m_{l}}} H(m_{l-1},s_{m_{l-1}},\dots,0,0)=0$.

Taking this relations and equation (\ref{spin}) into account, we obtain:
\begin{eqnarray}
&&\Delta^{(s_{m_{l}})} (\sigma^-) H(m_{l},s_{m_{l}},\dots,0,0)= \label{lab} \\
&&=\{ [\Delta^{(s_{m_{l}}-1)} (\sigma^+)]^{m_{l}-m_{l-1}-1} 
[\alpha_1- \alpha_0 (m_{l}-m_{l-1})(m_{l}+m_{l-1}-s_{m_{l}})]- \nonumber \\
&&-\alpha_1 \sigma^+_{(s_{m_{l}})} [\Delta^{(s_{m_{l}}-1)} 
(\sigma^+)]^{m_{l}-m_{l-1}-2} [(m_{l}-m_{l-1}-1)
(m_{l}+m_{l-1}-s_{m_{l}}-1)\} \nonumber \\
&& H(m_{l-1},s_{m_{l-1}},\dots,0,0) 
\end{eqnarray}   
The preceding expression can vanish (provided $\alpha_1 \neq 0$)
only if one of the following conditions is satisfied: 
\begin{eqnarray}
&& m_{l-1}=m_{l}-1 \label{sol1} \\
&& m_{l}+m_{l-1}=s_{m_{l}}+1 \label{sol2}
\end{eqnarray}
The second solution is not acceptable because of the inequality
(\ref{inequality}).

Substituting (\ref{sol1}) inside (\ref{lab}) and requiring it to be zero, we
obtain that the coefficients $\alpha_{0},\alpha_{1}$ must satisfy the equation:
\begin{equation}
\alpha_1=\alpha_0(2m_{l}-s_{m_{l}}-1) \label{rec3}
\end{equation}
We underline that this result could have been obtained directly from 
equation (\ref{rec2}) imposing $m_{l-1}=m_{l}-1$, $\lambda_{min}=-1$ and
$\alpha_i=a_{m_l-m_{l-1}-1}$ (we interchanged the indices in
(\ref{kernel}) with respect to (\ref{rec2})). Moreover, since $m_{0}=0$,
equation (\ref{sol1}) determines all the values  of the $m_{i}$,
$i=1,\dots,l$, so that we need to specify only the  value of the last one:
$m_{l}$. So we rename $m_{l}$ as $m$ and we  rewrite the eigenstates in the
more compact form: 
\begin{eqnarray} 
\phi(k,m,s_m,s_{m-1},\dots,s_1,0)=[
\Delta^{(N)}(\sigma^+)]^{k-m}  H(m,s_m,s_{m-1},\dots,s_1,0) \label{eigen2} \\
N \geq s_m \geq 2m \qquad s_m>s_{m-1} \qquad m \leq k \leq N-m 
\nonumber
\end{eqnarray}
\begin{eqnarray}
&& H(m,s_m,s_{m-1},\dots,s_1,0)= \nonumber \\ 
&&=\left( \sum_{i=0}^1 \alpha_i (\sigma^+_{(s_m)})^i
[\Delta^{(s_m-1)}(\sigma^+)]^{1-i} \right) H(m-1,s_{m-1},\dots,s_1,0)
\label{kernel2} \\
&& \alpha_1=\alpha_0(2m-s_m-1) \nonumber \\
&& H^{(0)}_0=\overbrace{|\downarrow \rangle \cdots |\downarrow \rangle}^N
\nonumber
\end{eqnarray}
We have shown that the states $H(m,s_m,\dots,0)$ belong to the kernel of 
$\Delta^{(N)} (\sigma^-)$; the proof that they form a basis for such space
is given in appendix A.

Now we want to show that the states (\ref{eigenstates}) are 
simultaneous eigenstates of the set of observables (\ref{family2}).
First of all it is easily seen that:
\begin{equation}
\Delta^{(N)}(\sigma^3)\phi(k,m,s_m,\dots,0)=
(2k-N)\phi(k,m,s_m,\dots,0) \label{value1}
\end{equation}
so let us consider the action of a partial Casimir $C_{l}$ on a state 
of the form (\ref{eigen2}). Since all the partial Casimirs commutes with
the $N$-th coproduct of any generator (\ref{algebra2}) we must 
only worry about the action of the partial Casimirs on the states
(\ref{kernel2}):
\begin{eqnarray*}
&& C_{l}H(m,s_m,\dots,0)=
\left( [\Delta^{(l)}(\sigma^3)]^2+4 \Delta^{(l)}(\sigma^+) \Delta^{(l)}(\sigma^-)
-2 \Delta^{(l)}(\sigma^3) \right)  \\
&& \left( \sum_{i=0}^1 \alpha_i (\sigma^+_{(s_m)})^i
[\Delta^{(s_m-1)}(\sigma^+)]^{1-i} \right) H(m-1,s_{m-1},\dots,s_1,0)
\end{eqnarray*}
There are two possibilities: if $l \geq s_{m}$ then we know that 
$$\Delta^{(l)}(\sigma^-)H(m,s_m,\dots,0)=0$$
and using equation (\ref{spin})
we obtain:
\begin{displaymath}
C_{l}H(m,s_m,\dots,0)=(2m-l)(2m-l-2)H(m,s_m,\dots,0)
\end{displaymath}
Viceversa, if $l<s_{m}$ then the Casimir operator $C_{l}$ commutes 
with both $\sigma^+_{(s_m)}$ and $\Delta^{(s_m-1)}(\sigma^+)$, so 
that we have:
\begin{eqnarray*}
&& C_{l}H(m,s_m,\dots,0)=\\
&& \left( \sum_{i=0}^1 \alpha_i (\sigma^+_{(s_m)})^i
[\Delta^{(s_m-1)}(\sigma^+)]^{1-i} \right) C_{l} H(m-1,s_{m-1},\dots,s_1,0)
\end{eqnarray*}
Again we will have two cases: if $l \geq s_{m-1}$ then it will hold
\begin{displaymath}
C_{l}H(m-1,s_{m-1},\dots,0)=(2m-l-2)(2m-l-4)H(m-1,s_{m-1},\dots,0)
\end{displaymath}
else $C_{l}$ will act directly on the state $H(m-2,s_{m-2},\dots,0)$.
We can iterate this procedure until we find a state $H(i,s_{i},\dots,0)$
such that $l \geq s_{i}$ (this will always be the case because 
$s_{0}=0$). Then using equations $\Delta^{(l)}(\sigma^-)H(i,s_i,\dots,0)=0$ 
and (\ref{spin}), we obtain:
\begin{displaymath}
C_{l}H(i,s_i,\dots,0)=(2i-l)(2i-l-2)H(m,s_m,\dots,0)
\end{displaymath}
Summarizing, we have:
\begin{equation}
C_l \phi(k,m,s_m,\dots,0)=(2i-l)(2i-l-2)\phi(k,m,s_m,\dots,0) \label{value2}
\end{equation}
where the value of $i$ is selected by the condition:
\begin{equation}
s_i \leq l <s_{i+1} \qquad s_{m+1}=N+1 \label{ineq2}
\end{equation}

Clearly, we can label the states (\ref{eigen2}) using the set of their
eigenvalues  under the complete set of observables (\ref{family2}) instead of
using the set of numbers $k,m,s_m,s_{m-1},\dots,0$. From equation
(\ref{value1}), we see that $k$ is common to both sets, so that we need to
take care only of the quantum numbers corresponding to the partial
Casimirs. From equation (\ref{value2}), we see that the eigenvalue
corresponding to $C_l$ is uniquely determined by the knowledge of   the
corresponding value of $i$; we will denote such value by $i_l$.  So we will
associate to a state (\ref{kernel2}) the set of $N-1$ numbers 
$i_2,\dots,i_N$. To the inequality (\ref{inequality}) it corresponds the
inequality  \begin{equation}
0 \leq i_l \leq [l/2] \label{i1}
\end{equation}
and from condition (\ref{ineq2}) we obtain the further
inequality: 
\begin{equation}
i_l \leq i_{l+1} \leq i_l +1  \quad l=2,\dots,N-1 \label{i2}
\end{equation}
To each set of numbers $i_2,\dots,i_N$ satisfying inequalities (\ref{i1}),
(\ref{i2}) it will correspond a state of the form (\ref{kernel}), so that there
exists a one to one mapping
\begin{equation}
\{ i_2,\dots, i_N \} \longleftrightarrow \{ m,s_m,s_{m-1},\dots,s_1,0 \}
\label{correspondence}
\end{equation}
Indeed, first of all we notice that since $s_m \leq N < s_{m+1}=N+1$ is always
verified, it  holds $i_N=m$. Moreover from condition (\ref{ineq2}) it follows
that if  $i_l-i_{l-1}=1$, then for that value of $l$ we have $s_{i_l}=l$.
Hence, the mapping (\ref{correspondence}) is given by:
\begin{displaymath}
\{ i_2,\dots, i_N \} \longrightarrow \{ i_N, \{s_{i_l}=l, \quad 2 \leq l \leq N,
\quad l \, \, s.t. \, \, i_l-i_{l-1}=1 \} ,0 \}
\end{displaymath}       

\section{Quantum deformed case}
\setcounter{equation}{0}
We briefly describe the results in the quantum deformed case. To this aim we
must pass from the UEA $U(sl(2))$ to its $q$-deformed version $U_q(sl(2))$
that is well known from the literature (see for example \cite{Hopf1}); the
generators of  $U_q(sl(2))$ satisfy the following commutations relations:
\begin{eqnarray*}
\left[ \sigma^{3}, \sigma^{+} \right]&=& 2 \sigma^+ \\
\left[ \sigma^{3}, \sigma^{-} \right]&=& -2 \sigma^{+} \\
\left[ \sigma^+, \sigma^- \right]&=& \frac{\sinh(z \sigma^3)}{\sinh z} 
\end{eqnarray*}
and an admissible co-product is defined by:
\begin{eqnarray*}
\Delta_z(\sigma^3)&=& \sigma^3 \otimes 1 + 1 \otimes \sigma^3 \\
\Delta_z(\sigma^+)&=& \sigma^+ \otimes e^{ \frac{z \sigma^3}{2}}  
+  e^{ -\frac{z \sigma^3}{2}} \otimes  \sigma^+ \\
\Delta_z(\sigma^-)&=& \sigma^- \otimes e^{ \frac{z \sigma^3}{2}}  
+  e^{ -\frac{z \sigma^3}{2}} \otimes  \sigma^-
\end{eqnarray*}
We can again define the extended co-products $\Delta_z^{(i)}$ through the
recursive formula (\ref{Deltarec}).
The Casimir operator for $U_q(sl(2))$ reads:
\begin{displaymath}
C_z=2 \left( 
\frac{\sinh [z (\sigma^3 -1)/2]}{\sinh z} \right)^2 +2 
\sigma^+ \sigma^-
\end{displaymath}
Now we have all we need to construct the $q$-deformed Gaudin magnet.
We consider the family of $N$ commuting observables:
\begin{eqnarray}
&& C_z^{(m)}=2 \left( 
\frac{\sinh [z ( \Delta_z^{(m)}(\sigma^3) -1)/2]}{\sinh z} \right)^2 +2 
\Delta^{(m)}(\sigma^+) \Delta^{(m)}(\sigma^-) \label{f1}\\
&& \qquad m=2,\dots,N  \nonumber \\
&& \Delta_z^{(N)}(\sigma^3) \label{f2}
\end{eqnarray}
The Casimir operator $C_z^{(N)}$ is the $q$-deformed version of the Gaudin
Hamiltonian (\ref{HG}). If we write down in detail the co-products inside
$C_z^{(N)}$ it takes the form:
\begin{displaymath}
C_z^{(N)}=\sum_{k,l=1}^N \left( \sigma^3_k \sigma^3_l J_{kl}^3 +
\sigma^+_k \sigma_l^- \tilde{J}_{kl} \right) + \sum_{k=1}^N  \sigma^3_k M^3_k
+ \frac{1}{2 \cosh^2(z/2)} e^{-z \Delta_z^{(N)}(\sigma^3)}
\end{displaymath} 
where the operators $J_{kl}^3 ,\tilde{J}_{kl}, M^3_k$ contain the non-local
part of the interaction and are given by:
\begin{eqnarray*}
&& J_{kl}^3= \frac{\cosh z}{\cosh^2(z/2)} e^{\frac{z}{2} \sum_{j=1}^N [
{\mathrm{sign}}(j-k)+ {\mathrm{sign}}(j-l)) ] \sigma^3_j}\\
&& \tilde{J}_{kl}=e^{\frac{z}{2} \sum_{j=1}^N [
{\mathrm{sign}}(j-k)+ {\mathrm{sign}}(j-l)) ] \sigma^3_j}
\left( 1+2{\mathrm{sign}}(k-l)  \sinh(\frac{z}{2}) \times \right. \\
&& \times \left. \left[ e^{\frac{z}{2} {\mathrm{sign}}(k-l)\sigma^3_l} +
e^{\frac{z}{2} {\mathrm{sign}}(l-k) \sigma^3_k)} \right] \right) \\
&& M^3_k= \left( \frac{\sinh(z/2)}{\cosh^2(z/2)} e^{-\frac{z}{2}
\Delta_z^{(N)}(\sigma^3)} \right) e^{\frac{z}{2} \sum_{j=1}^N [
{\mathrm{sign}}(j-k) \sigma^3_j ] } 
\end{eqnarray*}
All the results obtained in the undeformed case can be easily generalized to
the deformed one. In particular the complete family of observables (\ref{f1}),
(\ref{f2}) is simultaneously diagonalized by the ``deformed'' eigenstates:
\begin{eqnarray*} 
&& \phi_z(k,m,s_m,s_{m-1},\dots,s_1,0)=[
\Delta_z^{(N)}(\sigma^+)]^{k-m}  H_z(m,s_m,s_{m-1},\dots,s_1,0) \\ 
&& N \geq s_m \geq 2m \qquad s_m>s_{m-1} \qquad m \leq k \leq N-m \\
&& H_z(m,s_m,s_{m-1},\dots,s_1,0)= \\ 
&&=\left( \sum_{i=0}^1 \alpha_i(z) (\sigma^+_{(s_m)})^i
[\Delta_z^{(s_m-1)}(\sigma^+)]^{1-i} \right) H_z(m-1,s_{m-1},\dots,s_1,0) \\
&& \alpha_1(z) =\alpha_0(z) e^{\frac{z}{2}(2m-s_m-2)}
\frac{\sinh[z(2m-s_m-1)]}{\sinh z} \\ 
&& H^{(0)}_0=\overbrace{|\downarrow
\rangle \cdots |\downarrow \rangle}^N 
\end{eqnarray*} 
The corresponding eigenvalues are given by formulae:
\begin{eqnarray*}
&& \Delta^{(N)}(\sigma^3)\phi_z(k,m,s_m,\dots,0)=
(2k-N)\phi_z(k,m,s_m,\dots,0)\\
&& C_z^{(l)} \phi(k,m,s_m,\dots,0)=2 \left(
\frac{\sinh[z(2i-l-1)/2]}{\sinh z} \right)^2 \phi_z(k,m,s_m,\dots,0)
\end{eqnarray*} 
where the value of $i$ is selected by the
condition: 
\begin{displaymath}
s_i \leq l <s_{i+1} \qquad s_{m+1}=N+1 
\end{displaymath}

\section{Concluding remarks}

To conclude this paper, it might be worth noticing
 that what has been done for the 
spin $1/2$ representation can be extended with the proper obvious
modifications to any finite-dimensional representation. 
On the other hand, the transition to higher rank Lie algebras and
their $q$-deformation is nontrivial. In fact, as explained
in \cite{Balrag}, it turns out that for higher rank Lie algebras complete
integrability cannot be
taken for granted, neither at the classical nor at the quantum level: there
is a sensitive dependence on the representation, and the role of the so-called
``multiplicity free'' representations is crucial \cite{Leahy}.

A further possible generalization has to do with the Lie superalgebra 
framework \cite{Sorba}. Actually, work is in progress in this direction,
and we are looking for a supersymmetric integrable generalization of the 
Calogero-Gaudin system and of its $q$-deformation.

\newappendix{}
We know that:
\begin{equation}
dim[ker(\Delta^{(N)} (\sigma^-))]=\left( 
\begin{array}{c} 
N\\
\left[ \frac{N}{2} \right] 
\end{array} 
\right)
\label{dimension}
\end{equation}  
How many states of the form (\ref{kernel}) can we construct? 
If we fix $m$ and $s_m$ ($s_m \geq 2m$) we can construct a different state
for each state $H(m-1,s_{m-1},\dots,0)$ with $2(m-1) \leq s_{m-1}<s_{m}$. 
We will denote the total number of such states with $h(m,s_m)$, hence
\begin{displaymath}
\begin{array}{l}
h(m,s_m)=\sum_{s_{m-1}=2m-2}^{s_m-1} h(m-1,s_m-1) \qquad if s_m \geq 2m\\
h(m,s_m)=0 \qquad if \ s_m<2m\\
h(1,s_1)=1 \qquad if \  2 \leq s \leq N
\end{array}
\end{displaymath}
The solution of this series is given by:
\begin{equation}
h(m,s_m)=\left(
\begin{array}{c}
s_m\\
m-1
\end{array}
\right) - 2 \left(
\begin{array}{c}
s_m-1\\
m-2
\end{array}
\right) \label{soluzione}
\end{equation}
In fact it holds:
\begin{eqnarray*}
h(2,s_2)=\sum_{s_1=2}^{s_2-1} h(1,s_1)=\sum_{s_1=2}^{s_2-1} 1=
\left(
\begin{array}{c}
s_2\\
1
\end{array}
\right)- 2 \left(
\begin{array}{c}
s_2-1\\
0
\end{array}
\right)
\end{eqnarray*}
Moreover, equation (\ref{soluzione}) is satisfied; in fact: 
\begin{eqnarray*}
&& \sum_{s_{m-1}=2m-2}^{s_m-1}
\left(
\begin{array}{c}
s_{m-1}\\
m-2
\end{array}
\right)- 2 \left(
\begin{array}{c}
s_{m-1}-1\\
m-3
\end{array}
\right)=\\
&&=\sum_{s_{m-1}=2m-2}^{s_m-1} \left(
\begin{array}{c}
s_{m-1}\\
m-2
\end{array}
\right)- 2
\sum_{s_{m-1}=2m-3}^{s_m-2}\left(
\begin{array}{c}
s_{m-1}\\
m-3
\end{array}
\right)=\\
&& = \left(
\begin{array}{c}
s_{m}\\
m-1
\end{array}
\right)-\left(
\begin{array}{c}
2m-2\\
m-1
\end{array}
\right)-2 \left(
\begin{array}{c}
s_{m}-1\\
m-2
\end{array}
\right)+\left(
\begin{array}{c}
2m-3\\
m-2
\end{array}
\right)=\\
&&= \left(
\begin{array}{c}
s_m\\
m-1
\end{array}
\right)- 2 \left(
\begin{array}{c}
s_m-1\\
m-2
\end{array}
\right)
\end{eqnarray*} 
Where we used the equation \cite{librus}:
\begin{equation}
\sum_{k=m}^n \left(
\begin{array}{c}
k\\
a
\end{array}
\right) =
\left(
\begin{array}{c}
n+1\\
a+1
\end{array}
\right)-
\left(
\begin{array}{c}
m\\
a+1
\end{array}
\right) \label{librus}
\end{equation} 
Now, to obtain the total number of states of the form (\ref{kernel}) we must
sum $h(m,s)$ on all possible values for $m$ and $s$. Hence we have:
\begin{equation}
h_T=\left[ \sum_{m=1}^{[\frac{N}{2}]} \sum_{s=2m}^N h(m,s) \right] +1=
\left[ \sum_{m=1}^{[\frac{N}{2}]}
\left(
\begin{array}{c}
N+1\\
m
\end{array}
\right)-2
\left(
\begin{array}{c}
N\\
m-1
\end{array}
\right) \right]+1 \label{serie}
\end{equation}  
We have the following useful equations \cite{librus}:
\begin{eqnarray*}
&& \sum_{k=0}^{[ \frac{N}{2} ]}
\left(
\begin{array}{c}
N\\
k
\end{array}
\right) = 
2^{N-1} + \frac{1+(-1)^N}{4} \left(
\begin{array}{c}
N\\
\left[ \frac{N}{2} \right]
\end{array}
\right) \\
&& \sum_{k=0}^{[ \frac{N-1}{2} ]}
\left(
\begin{array}{c}
N\\
k
\end{array}
\right) = 2^{N-1}- \frac{1+(-1)^N}{4} \left(
\begin{array}{c}
N\\
\left[ \frac{N}{2} \right]
\end{array}
\right) 
\end{eqnarray*}
Hence
\begin{eqnarray*}
h_T&=&\sum_{m=1}^{[\frac{N+1-1}{2}]}
\left(
\begin{array}{c}
N+1\\
m
\end{array}
\right)
-2 \sum_{m=0}^{[\frac{N}{2}]-1}
\left(
\begin{array}{c}
N\\
m
\end{array}
\right)+1=\\
&=& 2 \left(
\begin{array}{c}
N\\
\left[ \frac{N}{2} \right]
\end{array}
\right) 
- \frac{1-(-1)^N}{4} \left(
\begin{array}{c}
N+1\\
\left[ \frac{N+1}{2} \right]
\end{array}
\right)- \frac{1+(-1)^N}{4} \left(
\begin{array}{c}
N\\
\left[ \frac{N}{2} \right]
\end{array}
\right)
\end{eqnarray*}
In both the case when $N$ is even and when $N$ is odd we obtain the same result, namely:
\begin{displaymath}
h_T= \left(
\begin{array}{c}
N\\
\left[ \frac{N}{2} \right]
\end{array}
\right) 
\end{displaymath}
that is exactly the dimension of the kernel of the operator $\Delta^{(N)}(\sigma^-)$.

Now we want to show that the states (\ref{eigenstates}) form a basis in the
Hilbert space of the problem, i.e., that the total number of such states is
$2^N$. 

We know that for a given value of $m$, $k$ can assume
$N-2m+1$ different values. The total number of states of the form 
(\ref{eigenstates}) is then given by ($h(m)=\sum_{s=2m}^N h(m,s)$)
\begin{eqnarray*}
&& \left[ \sum_{m=1}^{[\frac{N}{2}]} (N-2m+1) h(m) \right] +N=\\
&&= (N+1) \left(
\begin{array}{c}
N\\
\left[ \frac{N}{2} \right]
\end{array}
\right)
-2 \sum_{m=1}^{[\frac{N}{2}]} m \left(
\begin{array}{c}
N+1\\
m
\end{array}
\right)+4 \sum_{m=1}^{[\frac{N}{2}]} m
\left(
\begin{array}{c}
N\\
m-1
\end{array}
\right)=\\
&&=(N+1) \left(
\begin{array}{c}
N\\
\left[ \frac{N}{2} \right]
\end{array}
\right)
-2(N+1) \sum_{m=1}^{[\frac{N}{2}]} \left( 
\begin{array}{c}
N\\
m-1
\end{array}
\right) +\\
&&+ 4N \sum_{m=0}^{[\frac{N-2}{2}]-1} \left( 
\begin{array}{c}
N-1\\
m
\end{array}
\right)+4 \sum_{m=0}^{[\frac{N}{2}]-1} \left(
\begin{array}{c}
N\\
m
\end{array}
\right)=\\
&&=(3N-1) \left(
\begin{array}{c}
N\\
\left[ \frac{N}{2} \right]
\end{array}
\right)-[1+(-1)^N] \left( \frac{N-1}{2} \right) \left(
\begin{array}{c}
N\\
\left[ \frac{N}{2} \right]
\end{array}
\right)+\\
&&+2^N - [1+(-1)^N-1] N  \left( 
\begin{array}{c}
N-1\\
\left[ \frac{N-1}{2} \right]
\end{array}
\right)-4N \left(
\begin{array}{c}
N-1\\
\left[ \frac{N}{2} \right]-1
\end{array}
\right) 
\end{eqnarray*}
In both the cases when $N$ is even or odd, this expression yield the desired result:$2^N$.    

\newappendix{}
We give the explicit form of the states (\ref{kernel}) in the case $N=5$
and for the choice of the normalization parameter $\alpha_0=1$:
\begin{eqnarray*}
H(0,0)&=&|\downarrow \downarrow \downarrow \downarrow \downarrow \rangle\\
H(1,2)&=&|\uparrow \downarrow \downarrow \downarrow \downarrow \rangle -
|\downarrow \uparrow \downarrow \downarrow \downarrow \rangle\\
H(1,3)&=&|\uparrow \downarrow \downarrow \downarrow \downarrow \rangle +
|\downarrow \uparrow \downarrow \downarrow \downarrow \rangle -
2 |\downarrow \downarrow \uparrow \downarrow \downarrow \rangle\\
H(1,4)&=&|\uparrow \downarrow \downarrow \downarrow \downarrow \rangle +
|\downarrow \uparrow \downarrow \downarrow \downarrow \rangle +
|\downarrow \downarrow \uparrow \downarrow \downarrow \rangle
-3|\downarrow \downarrow \downarrow \uparrow \downarrow \rangle\\
H(1,5)&=&|\uparrow \downarrow \downarrow \downarrow \downarrow \rangle +
|\downarrow \uparrow \downarrow \downarrow \downarrow \rangle +
|\downarrow \downarrow \uparrow \downarrow \downarrow \rangle +
|\downarrow \downarrow \downarrow \uparrow \downarrow \rangle
-4 |\downarrow \downarrow \downarrow \downarrow \uparrow \rangle\\
H(2,4,2)&=&|\uparrow \downarrow \uparrow \downarrow \downarrow \rangle +
|\downarrow \uparrow \downarrow \uparrow \downarrow \rangle -
|\downarrow \uparrow \uparrow \downarrow \downarrow \rangle -
|\uparrow \downarrow \downarrow \uparrow \downarrow \rangle\\
H(2,4,3)&=& 2 |\uparrow \uparrow \downarrow \downarrow \downarrow \rangle +
2 |\downarrow \downarrow \uparrow \uparrow \downarrow \rangle -
|\uparrow \downarrow \uparrow \downarrow \downarrow \rangle -
|\downarrow \uparrow \downarrow \uparrow \downarrow \rangle -
|\downarrow \uparrow \uparrow \downarrow \downarrow \rangle -\\
& & - |\uparrow \downarrow \downarrow \uparrow \downarrow \rangle\\
H(2,5,2)&=& 2 |\downarrow \uparrow \downarrow \downarrow \uparrow \rangle +
|\uparrow \downarrow \uparrow \downarrow \downarrow \rangle +
|\uparrow \downarrow \downarrow \uparrow \downarrow \rangle -
|\downarrow \uparrow \uparrow \downarrow \downarrow \rangle -
|\downarrow \uparrow \downarrow \uparrow \downarrow \rangle -\\
& & - 2 |\uparrow \downarrow \downarrow \downarrow \uparrow \rangle\\
H(2,5,3)&=& 2|\uparrow \uparrow \downarrow \downarrow \downarrow \rangle-
|\uparrow \downarrow \uparrow \downarrow \downarrow \rangle +
|\uparrow \downarrow \downarrow \uparrow \downarrow \rangle -
|\downarrow \uparrow \uparrow \downarrow \downarrow \rangle +
|\downarrow \uparrow \downarrow \uparrow \downarrow \rangle -\\
& & - 2 |\downarrow \downarrow \uparrow \uparrow \downarrow \rangle -
2 |\uparrow \downarrow \downarrow \downarrow \uparrow \rangle -
2 |\downarrow \uparrow \downarrow \downarrow \uparrow \rangle +
4|\downarrow \downarrow \uparrow \downarrow \uparrow \rangle \\
H(2,5,4)&=& 2|\uparrow \uparrow \downarrow \downarrow \downarrow \rangle +
2 |\uparrow \downarrow \uparrow \downarrow \downarrow \rangle -
2 |\uparrow \downarrow \downarrow \uparrow \downarrow \rangle +
2 |\downarrow \uparrow \uparrow \downarrow \downarrow \rangle -
2 |\downarrow \downarrow \uparrow \uparrow \downarrow \rangle -\\
& & - 2 |\uparrow \downarrow \downarrow \downarrow \uparrow \rangle -
2 |\downarrow \uparrow \downarrow \downarrow \uparrow \rangle -
2 |\downarrow \downarrow \uparrow \downarrow \uparrow \rangle -
2 |\downarrow \uparrow \downarrow \uparrow \downarrow \rangle +
6 |\downarrow \downarrow \downarrow \uparrow \uparrow \rangle
\end{eqnarray*}

\end{document}